\documentclass[apj]{emulateapj}

\usepackage{natbib}
\usepackage{graphicx}
\bibpunct{(}{)}{;}{a}{}{,}

\begin{document}

   \title{
	Convectively driven vortex flows in the Sun}

   \author{
 	J.~A.~Bonet\altaffilmark{1},
  	I.~M\'arquez\altaffilmark{1,2},
   	J.~S\'anchez~Almeida\altaffilmark{1},
	I.~Cabello\altaffilmark{3},
	and
	V.~Domingo\altaffilmark{3}
	}
    \altaffiltext{1}{Instituto de Astrof\'\i sica de Canarias, 
              E-38205 La Laguna, Tenerife, Spain}
    \altaffiltext{2}{Departamento de An\'alisis
     	Matem\'atico, Universidad de La Laguna, E-38271 La Laguna, 
     	Tenerife, Spain}
    \altaffiltext{3}{Grupo de Astronom\'\i a y Ciencias del Espacio, 
	Universidad de Valencia, E-46980 Paterna, Valencia, Spain}
   \email{jab@iac.es, imr@iac.es, jos@iac.es, iballa.cabello@uv.es,
	vdomingo@uv.es}

\begin{abstract}
We have discovered small whirlpools in the Sun, with a size 
similar to the terrestrial hurricanes ($\la$~0.5~Mm). 
The theory of solar convection predicts them, 
but they had remained elusive so far. The vortex flows are 
created at the downdrafts where the plasma returns to 
the solar interior after cooling down, and we detect them 
because some magnetic bright points (BPs) follow a logarithmic
spiral in their way to be engulfed by a downdraft.
Our disk center observations show 0.9$\times 10^{-2}$~vortexes per Mm$^{2}$, 
with a lifetime of the order of 5~min, and with
no preferred sense of rotation. They are not 
evenly spread out over the surface, but they seem to
trace the supergranulation and the mesogranulation. 
These observed properties are strongly biased 
by our type of 
measurement, unable to detect vortexes 
except when they  are engulfing magnetic BPs.
\end{abstract}

   \keywords{
	convection -- Sun: photosphere -- Sun: granulation
               }


\shorttitle{Convectively driven vortex flows in the Sun}
\shortauthors{Bonet et al.}

\section{Introduction}\label{introduction}

A proper 
understanding of the solar surface convection  has 
impact on many different
areas of astrophysics. For example, (a) the solar metallicity 
is used as a universal reference, and it has been 
recently modified by fifty percent after considering 
realistic convective motions in the analysis \citep{asp05b},
and (b) 
turbulent dynamo action can produce magnetic
fields in many astrophysical environments
\citep[e.g.,][]{mar04,ber05},
but its observational study is so far confined to the 
turbulent dynamo driven by the solar convective motions
\citep[e.g.,][]{cat99a,vog07}.

We have discovered small whirlpools in 
the Sun, with a size similar to the terrestrial 
hurricanes. This discovery confirms a specific
prediction of the current theory of solar convection
that had remained elusive so far.
The energy produced in the solar interior is 
transported to the surface firstly by radiation exchange 
and, during the last thirty percent of the way, by 
convective motions \citep[e.g.,][]{sti91}. 
The hot buoyant plasma rises, releases energy, 
and then falls down. According to the 
current theory, these convective motions are driven 
by strong highly localized downdrafts at the solar 
surface \citep{spr90,stei98,ras98}. 
The downdrafts are sinks where 
the cold plasma returns to the solar interior. 
Since the matter has angular momentum with respect to 
the draining point, it must spin up when approaching the 
sink, giving rise to a whirl flow ({\em bathtub effect}). 
Although the vortex motions predicted by the theory
have been repeatedly sought, they had not been found 
so far \citep{nov89,spr90,rou97,hoe98,nis03}. (The exception in 
\citealt{bra88} represents 
the same phenomenon at much larger scale.) We have 
detected these vortexes as proper motions of magnetic 
bright points (BPs), which follow spiral paths in the way 
to be engulfed by a downdraft. The BPs are 
easily blurred by seeing \citep{tit96}, thus the unique 90~km spatial
resolution provided by the Swedish Solar Telescope at 
La Palma turned out to be necessary \citep{sch02}. 
Magnetic BPs are present almost
everywhere on the Sun, also
in the quiet internetwork regions \citep{san04a,dew05,dew08}.
Precisely looking for properties of these  quiet Sun 
BPs, we found out that some of them spiral when heading 
towards the supposed location of a
sink, which is the result reported in this Letter. 
We note that the vortical motion of photospheric
magnetic concentrations may have significant
impact on its own, e.g., on the heating of the solar corona
\citep[][]{bal98,gud05,ama08}. 
Previous searches for photospheric vortexes 
were often motivated by this interest 
\citep{bal98,nis03}.

\section{Observations}\label{reduction}
A quiet Sun internetwork target was observed near solar 
disk center with the Swedish Solar Telescope 
\citep[SST; ][]{sch02}. During observation seeing
was fluctuating with minutes of excellence 
(September 29, 2007; from 9:13:57 to 9:45:55 UT).
We used 12 bit Kodak Mega Plus II 
2048~$\times$~2048 cameras, which render
an image scale of 0.034~\arcsec~pix$^{-1}$.
The field-of-view (FOV)
includes various supergranules.  
We observe simultaneously in
the G-band (a 10.8~\AA\ wide filter centered at 4305.6~\AA),
in 
Ca~H line-core, 
and in the continuum close to  Ca~H.
The images were reconstructed with 
multi-frame blind deconvolution
\citep[][]{noo05} 
to produce movies whose snapshots 
have an angular resolution close to the diffraction
limit of the SST 
at the working wavelength ($\sim$0\farcs 1). 
Each snapshot results from combining 125 images,
rendering a mean cadence of one snapshot every  15~s.
A destretching algorithm allows us to
remove residual image distortion left by the 
restoration 
process,
and a subsonic filter suppresses 
p-modes and residual jitter stemming from destreching
\citep[][]{tit89}.
Our analysis is focused on the G-band images, 
where BPs are easily observed even in the 
quietest Sun.

\section{Whirlpools}\label{whirlpool}
The G-band movie often shows 
BPs swirling around intergranular points where several
dark lanes converge. These motions are reminiscent 
of the bathtub vortex flows predicted by  
numerical simulations of convection, and which are 
driven by the granulation downdrafts (\S~\ref{introduction}). 
Since we have no other information but the proper motions 
of the BPs,  we cannot proceed without assuming that 
these proper motions trace true plasma motions. 
We cannot think of any other physical mechanism able 
to generate these {\em very special} trajectories, but it must be 
clear that treating the proper motions as plasma motions is a 
working  hypothesis\footnote{One may argue that BPs are not
good passive tracers because they posses strong kG 
magnetic fields and, therefore, the magnetic forces 
back-react on the plasma flows counteracting the drag forces.
This theoretical prejudice has been proved to be erroneous,
since the BPs move exactly as dictated by the 
granular flow field \citep[see ][]{bal98}. The 
reason seems to be the small cross section of the 
magnetic concentrations. Drag forces 
overcome magnetic forces when the magnetic
concentrations are slender enough
\citep[e.g.,][]{sch86,san01b}.}.

\begin{figure}
\includegraphics[width=0.5\textwidth]{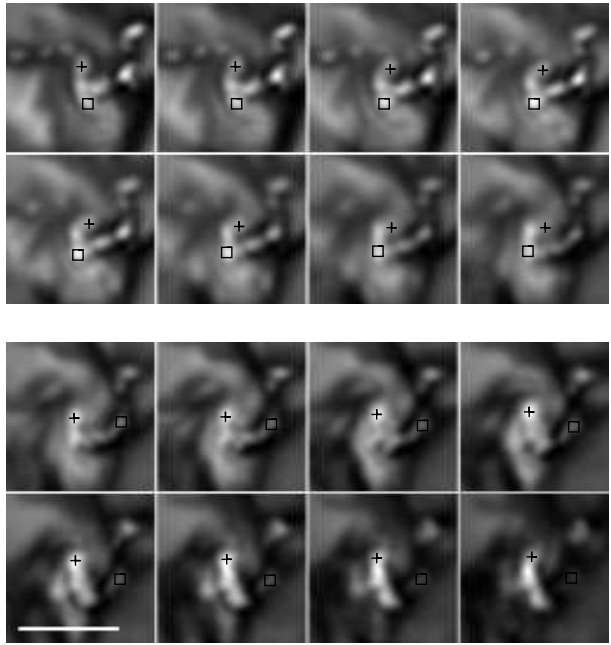}
\caption{Sequence of snapshots corresponding to
	the event analyzed in detail in the Letter. 
	The  snapshots are separated by 15~s, and the
	bar on the lower left image corresponds
	to 1~Mm on the Sun.
	We have
	marked with symbols two sets of two 
	nearby  BPs, so that it is easy  to appreciate how the two 
	BPs rotate with respect to each other.
	(Although the sequence is continuous, the two upper
	rows show only one pair, whereas the two lower rows
	show the other pair.)
	Note how the bright 
	structures in each snapshot depict a
	spiral pattern resembling 
	the vortex derived from BP proper motions.
	The 
	animation 
	given as on-line material allows
	to appreciate the vortex motions in 
	more detail.
	}
\label{snapshots}
\end{figure}

\begin{figure}
 \includegraphics[width=1.\textwidth]{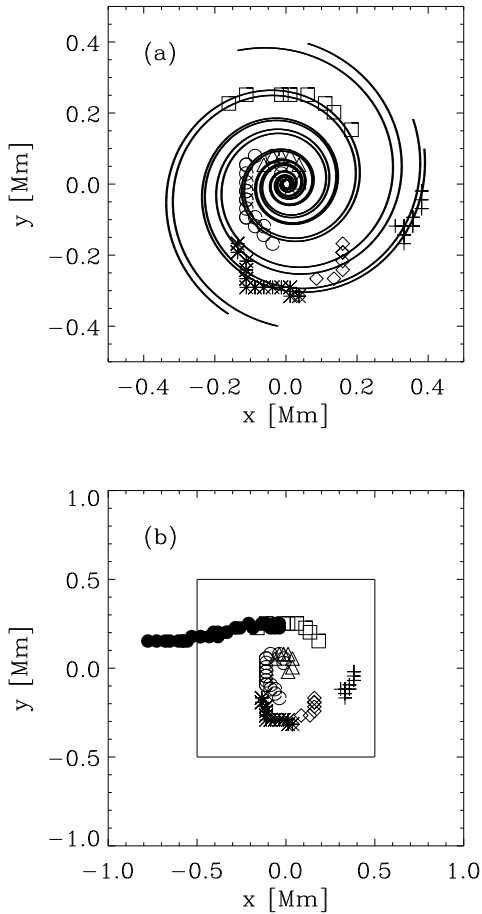}
\caption{
(a) Logarithmic spiral (the solid lines) that
fits the trajectories of six observed BPs (the symbols).
(b) Same as (a) but with a FOV twice larger and 
including two additional BPs (the solid circles)
whose paths are not set by the whirlpool, 
but they follow a nearby intergranular lane. 
The square in (b) represents the FOV in (a).
(The assignment of symbols to BPs 
differs from that used in Fig.~\ref{snapshots}.)
}
\label{todo}
\end{figure}

We will discuss statistical properties of the 
whirlpools in \S~\ref{statistics}. This section
is focused on a particularly well defined
event whose snapshots are shown in Fig.\ref{snapshots}.
(See also the animation provided 
as on-line material.) Up to eight different BPs move under 
the influence of this whirlpool during the 5~min
duration of the event. Their trajectories are shown
in Fig.~\ref{todo}, where we use a different symbol
for each trajectory, i.e., the set of positions
of a single BP as time goes on.
The positions have been determined as the 
centroid of the BP in successive snapshots.
The identification of each particular BP has been
carried out by eye, playing back and forth the movie
with the FOV of the event. It is clear from 
Fig.~\ref{todo}a that the BPs swirl around a
quite stable central point. Only two BPs follow
straight trajectories (see Fig.~\ref{todo}b),
but this path coincides with the position 
of one of the intergranular lanes that feeds
the whirlpool, and which channels the BP motion
(see also the top row of Fig.~\ref{snapshots}).  

If the analogy with a bathtub vortex holds for the
observed whirlpools, then a passive tracer should 
approximately follow a logarithmic spiral when heading
to the sinkhole, i.e., 
\begin{equation}
\theta-\theta_0=A\,\ln(r/r_0),
\label{bathtub}
\end{equation}
where $r$ and $\theta$ stand for
the radius and azimuth 
in polar coordinates 
with the draining point as origin.
The symbol $\theta_0$ corresponds to the azimuth at the 
arbitrary radius $r_0$, and 
\begin{equation}
A=U_\theta/U_r,
\label{bathtub2}
\end{equation}
is a constant of motion given by the ratio between
the azimuthal velocity $U_\theta$ and 
the radial velocity $U_r$. The parameter
$A$  controls the shape 
of the spiral, which 
becomes a straight line for $A$ equals zero, 
and a circle for $A\rightarrow\infty$.
Equation~(\ref{bathtub}) follows from 
the conservation of mass and angular momentum 
in a non-viscous incompressible fluid moving in a bathtub 
vortex \citep[e.g.,][]{fey64,mas83}. 
These conditions are only  loosely met by the solar plasma, 
however, the observed motions match quite well 
this approximate behavior.
Using a non-linear least squares procedure, we have fitted
the trajectories of the BPs in Fig.~\ref{todo}a. We assume 
the whirlpool to be the same for all of them,  i.e., 
all trajectories share a common origin and the same $A$. 
Each  BP trajectory has its own starting point 
parameterized by $\theta$ at $r_0$, i.e., by
$\theta_0$. The fitting is
fairly robust and independent of the
initial conditions, providing,
\begin{equation}
A=6.4 \pm 1.6,
\label{actuala}
\end{equation}
where the mean value corresponds to the fit in
Fig.~\ref{todo}a (the solid lines),
and the error bar indicates the standard 
deviation among the inferred values when 
each one of the 
BP trajectories is excluded from the fit.
The area of the swirling region turns out to be
of the order of 
0.5$\times$0.5~Mm$^2$. The size of the sinkhole 
(i.e., of the hypothetical downdraft) has to be smaller than 
the radius of curvature of the smallest trajectory, 
which corresponds to the one marked by triangles 
in Fig.~\ref{todo}. 
Its radius of curvature is of the order of 100 km, i.e., 
the downdraft seems to be  spatially unresolved to
SST observations. 
BP proper motion velocities span from one to four km~s$^{-1}$
and, with some noise, they tend to increase towards
the sinkhole. 

\section{Statistical properties}\label{statistics}
\begin{figure}
\includegraphics[width=0.45\textwidth]{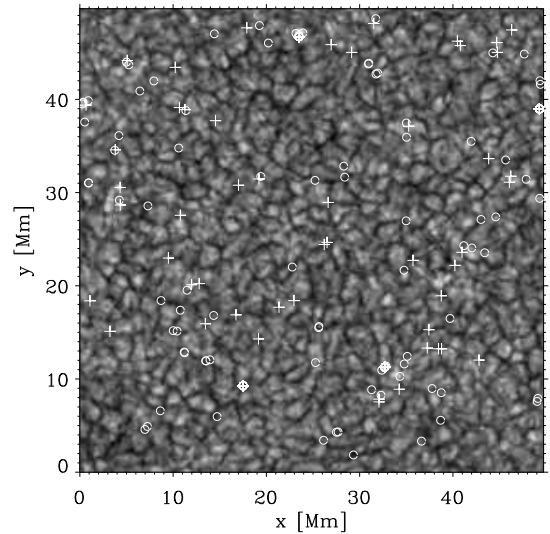}
\caption{
Positions of the observed
vortexes on the average G-band image.
They trace a pattern resembling the supergranulation
and the mesogranulation. Plus signs
correspond to counterclockwise motions whereas
circles correspond to clockwise motions. We do not
think that such a distribution reveals an intrinsic
property of the vortexes, but it is a bias
set by using magnetic BPs to trace small
scale proper motions.
}
\label{meso-super}
\end{figure}

Detailled visual inspection of the G-band
movie shows 
the whirlpool motion described above to be quite common.
In order to have a statistically significant
description of the phenomenon, we carried out
a careful visual analysis of the full time series looking 
for swirl motions.  Roughly speaking, one finds two
types of motions that we have identified as vortexes:
(1) two or more BPs move (in the same
or in opposite directions) towards a
point, where they rotate 
with respect to one another,
and (2) 
the rotation also includes 
the diffuse bright borders of the 
sometimes fragmented
granules 
surrounding the turning point. 
Taking this vague definition in
mind, we find 138 swirls (Fig.~\ref{meso-super}), 
83\% of which belong to type~1, and the rest to type~2.
(The motion shown in Fig.~\ref{snapshots} 
belongs to the second type.) Our FOV covers
49.7\,Mm$\times$49.7\,Mm. 
The time series lasts 31.5~min. 
This corresponds to  an space-and-time density 
equals to $d\simeq 1.8\times 10^{-3}$ vortex Mm$^{-2}$\,min$^{-1}$.

The lifetimes of the vortexes 
were also estimated. The starting time was defined 
as the time where the BPs begin their swirling motions. 
These motions often end up producing a central
still BP. 
Then we take the end time as the moment when this 
BP starts fading away or fragmenting. If the central 
BP is not formed,
our final time is the time when circular motions
are no longer distinguished. The mean lifetime 
of the 138 whirlpools turns out
to be $\tau\simeq (5.1\pm 2.1)~$min,
with the error bar representing the standard 
deviation among all single measurements.
Assuming that the number of swirls 
is stationary, at any given time one finds
$\tau d$ whirlpools on the solar surface,
which corresponds to 
0.90$\times 10^{-2}$ vortex Mm$^{-2}$.
The lifetime estimates are only lower limits to the
true lifetimes, since the vortexes should exist and 
last even if there are no magnetic 
BPs that allow us to detect them.
For the same reason the number of detected
vortexes also represents a lower limit. 

There is the same probability of clockwise and
counterclockwise motions. We find 50.4\% of the whirls
with counterclockwise rotation, and 49.3\% with clockwise 
rotation (Fig.~\ref{meso-super}).

The observed swirls are not uniformly distributed 
throughout the FOV. They follow a pattern that 
closely resembles supergranulation and mesogranulation
cells (Fig.~\ref{meso-super}).
It is not clear whether this distribution reveals 
an intrinsic property of the vortexes. Rather, it
seems to be a selection effect we impose when using 
G-band BPs as passive tracers of plasma motions. 
It is well known that BPs trace 
supergranulation \citep{dun73,meh74}
and mesogranulation \citep{san04a}.
Vortexes outside these areas
lack of enough BPs and they will not be detected.

\section{Discussion}

\begin{figure}
\includegraphics[width=0.5\textwidth]{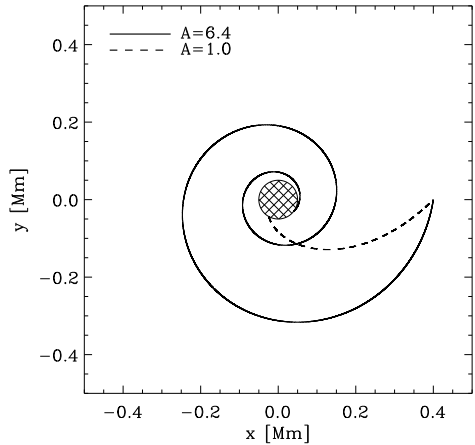}
\caption{
BP trajectories like the one analyzed in \S~\ref{whirlpool} 
(the solid line, corresponding to $A=6.4$), 
and what we expect to be the most probable case, with $A\sim 1$ 
(the dashed line).  The swirling motions of the latter are 
difficult to detect since the deviations from straight lines 
are of the order of the angular resolution of the 
observation (represented by the hashed disk 
at the core of the whirlpool).  
}
\label{twocases}
\end{figure}

Observing these convectively driven vortex flows 
has been elusive because of the size of the whirlpools. 
Motions take place at sub-arcsec scales  and, 
therefore, close to the best angular resolution 
achieved at present. In addition, the
case analyzed in  \S~\ref{whirlpool} seems to 
be extreme in the sense that the 
ratio between the azimuthal 
velocity and the radial velocity is
particularly large ($A\simeq 6.4$), which renders
closed spiral paths and clear swirling motions.
Assuming that the velocity of the 
plasma that feeds the whirlpools has an important 
stochastic component set by the complex 
granular dynamics, then one should expect 
the azimuthal and the radial velocities to be similar, 
rendering $A\sim 1$.
In this case the swirling motions are moderate 
and so easy to overlook.
The BP trajectories corresponding to these 
more typical
cases are hard to distinguish from straight lines within the angular 
resolution of the observations (see Fig.~\ref{twocases}).

Our work should be complemented with a 
spectroscopic confirmation of the 
vortex motions. It requires observing the whirlpools 
out of the disk center so that
the horizontal circular motions
have a component along the 
line-of-sight (LOS).
Such spectroscopic observation is challenging.
Achieving the required spatial resolution 
is even more difficult in spectroscopy.
Moreover, one would like to maximize
the velocity component along the LOS
by observing well outside 
the solar disk center. However, the solar surface 
is corrugated and, outside the disk center, 
the granules protrude hiding the 
intergranular lanes that harbor the vortexes 
\citep[e.g.,][]{car04,kel04}. 


\acknowledgements
We thank an anonymous referee for help with the
presentation of the results.
The work has partly been funded by the Spanish Ministry 
of Education, and Science and Innovation , projects 
AYA2007-66502, AYA2007-63881 and ESP2006-13030-C06-04.
Financial support by the European Commission through the SOLAIRE
Network (MTRN-CT-2006-035484) is gratefully acknowledged,
as are the computer resources and assistance
provided by the LaPalma (IAC/RES, Spain)
supercomputer installation used for image restoration.
{\it Facility:} \facility{SST}

%

%
%

\end{document}